# Controlling Surface-plasmon-polariton Launching with Hot Spot Cylindrical Waves in a Metallic Slit Structure

*Wenjie Yao, Chengwei Sun, Jianjun Chen*, and Qihuang Gong*

W. Yao
State Key Laboratory for Mesoscopic Physics and Department of Physics, Peking University, Beijing 100871, China
C. Sun, Prof. J. Chen, Prof. Q. Gong
State Key Laboratory for Mesoscopic Physics and Department of Physics, Peking University, Beijing 100871, China
Collaborative Innovation Center of Quantum Matter, Beijing, China
E-mail: jjchern@pku.edu.cn

Plasmonic nanostructures, which are used to generate surface plasmon polaritions (SPPs), always involve sharp corners where the charges can accumulate. This can result in strong localized electromagnetic fields at the metallic corners, forming hot spots. The influence of the hot spots on the propagating SPPs are investigated theoretically and experimentally in a metallic slit structure. It is found that the electromagnetic fields radiated from the hot spots, termed as the hot spot cylindrical wave (HSCW), can greatly manipulate the SPP launching in the slit structure. The physical mechanism behind the manipulation of the SPP launching with the HSCW is explicated by a semi-analytic model. By using the HSCW, unidirectional SPP launching is experimentally realized in an ultra-small metallic step-slit structure. The HSCW bridges the localized surface plasmons and the propagating surface plasmons in an integrated platform and thus may pave a new route to the design of plasmonic devices and circuits.

## 1. Introduction

Surface plasmon polaritons (SPPs) are electromagnetic waves that propagate along the metal-dielectric interface.[1] They are considered to be the promising candidate for the next-generation integrated photonic circuits, owing to the capability of confining the electromagnetic field below the diffraction limit.[2–6] Recent works have pointed out that SPPs are also likely to be the candidate carrier in future quantum information systems.[7–11] In order to realize such potential, it's crucial to launch the SPPs effectively and flexibly from the free-space light. For the widely used nanostructures, such as nanoslit,[12–18] nanogroove,[19,20] and nanoantenna,[21,22] all of them have sharp metallic corners where the charges can accumulate there. As a result, the localized electromagnetic fields become extremely strong at the sharp corners, which are usually termed as hot spots.[23–29] Because of the strong field enhancement, hot spots have been applied in enormous areas of the second harmonics, Raman enhancement, surface-enhanced optical sensing, spectroscopy, and microscopy.[23–28] However, seldom works have done to concern the conversion of the localized hot spots to the nearby propagating SPPs in a platform. This conversion between the localized hot spots and propagating SPPs will be an important issue in integrated plasmonic circuits, and it can also provide a new degree of freedom in designing SPP-based devices.

In this paper, the hot spots at the sharp metallic corners are investigated theoretically and experimentally in a metallic slit structure. It is found that the electromagnetic fields radiated from the hot spots can greatly manipulate the SPP launching in the slit structure. When a p-polarized light illuminates the slit structure, the charges accumulate at the slit corners, forming hot spots there. The electromagnetic fields radiated from the hot spots, termed as the hot spot cylindrical wave (HSCW), can convert to the SPPs by scattering and greatly influence the SPP launching. A semi-analytic model is established to explicate how the HSCW influences the propagating SPPs along the front metal surface. By using the HSCW, the SPP launching is manipulated, and unidirectional SPP launching with a high extinction ratio is achieved in an ultra-small metallic step-slit structure, which is successfully demonstrated in the experiment.

## 2. Numerical results and analysis

The investigated metallic slit structure is schematically shown in Figure 1(a), which is fabricated on a gold film. In the metallic slit structure, the left-side gold film is lower than the right one with the thickness difference of $d$. The thickness of the left-side gold film is $h$, and the slit width is $w$. In the case of $d = 0$ nm, the investigated structure is a conventional symmetric slit, which was widely used in plasmonic structures and devices.[12–15] In the case of $d > 0$ nm, the metallic slit becomes an asymmetric step-slit structure. For simplicity, the lower ($0 < y < h$) and upper ($h < y < h+d$) parts of the step-slit structure are named as the lower-slit and the upper-slit, respectively. In our analysis, the incident light source is set to be a p-polarized plane wave (magnetic vector parallel to the slit) with a uniform magnetic field along the down-port of the lower-slit, as indicated in Figure 1(a). The thickness of the left-side gold film is set to be $h = 200$ nm. The wavelength of the incident light is $\lambda = 830$ nm and the corresponding permittivity of the gold is $\varepsilon_{Au} = -16.61 + 1.665i$.[30] The simulations are performed by COMSOL MULTIPHYSICS throughout this paper. The SPP intensity and amplitude are obtained by using the mode orthogonality condition.[32] To begin with, the symmetric slit ($d = 0$) is considered. The intensities of the right- and left-propagating SPPs generated by the symmetric slit of different slit widths are depicted in Figure 1(b). It is noticed that the intensities of the right- and left-propagating SPPs along the front metal surface are the same for any slit widths because of the structural symmetry [the black line and the red line overlap completely with each other in Figure 1(b)]. Besides, synchronous oscillation curves for the right- and left-propagating SPPs with a period of $T = 830$ nm are obviously observed, and the oscillation amplitude becomes small as $w$ increases, as shown in Figure 1(b). This damping oscillation behavior has been observed in the previous works.[31, 32] However, the underlying physics of this phenomenon still remains unclear. Due to the different incident waves,[32] this damping oscillation behavior is a little different from that in the previous works.[32] Moreover, by increasing the thickness difference of $d$ to construct an asymmetric step-slit structure, the oscillation curves of the right- and left-propagating SPPs move oppositely, as shown in Figure 1(c)-1(e). For example, when $d$ is one quarter of the wavelength ($d = \lambda/4 = 207.5$ nm) [Figure 1(c)], the intensity of the right-propagating SPPs oscillates anti-synchronously in comparison with that of the left-propagating SPPs. When $d$ is half of the wavelength ($d = \lambda/2 = 415.0$ nm) [Figure 1(e)], the two curves return to synchronous oscillations. In analog with the oscillation of propagating waves, we define a relative phase difference of these two curves: in the case of synchronous oscillation curves, the relative phase difference between them is $\varphi = 0$ [Figure 1(b)]; and in the case of anti-synchronous oscillation curves, the relative phase difference is $\varphi = \pi$ [Figure 1(c)]. Therefore, the relative phase difference between the two oscillation curves approximately equals $\varphi = 2k_0 d$, where $k_0 = 2\pi/\lambda$ is the free-space wave vector. Near the relative phase difference of $\varphi = \pi$ ($d = 250$ nm), the left-propagating SPPs almost vanished at the slit width of $w = 500$ nm, while the right-propagating SPP intensity almost reaches a peak, as denoted by the blue dashed line in Figure 1(d). In this case, unidirectional SPP launching with a high extinction ratio of about 25 can be achieved. Here, the extinction ratio is defined as the quotient of the right- and left-propagating SPP intensities.

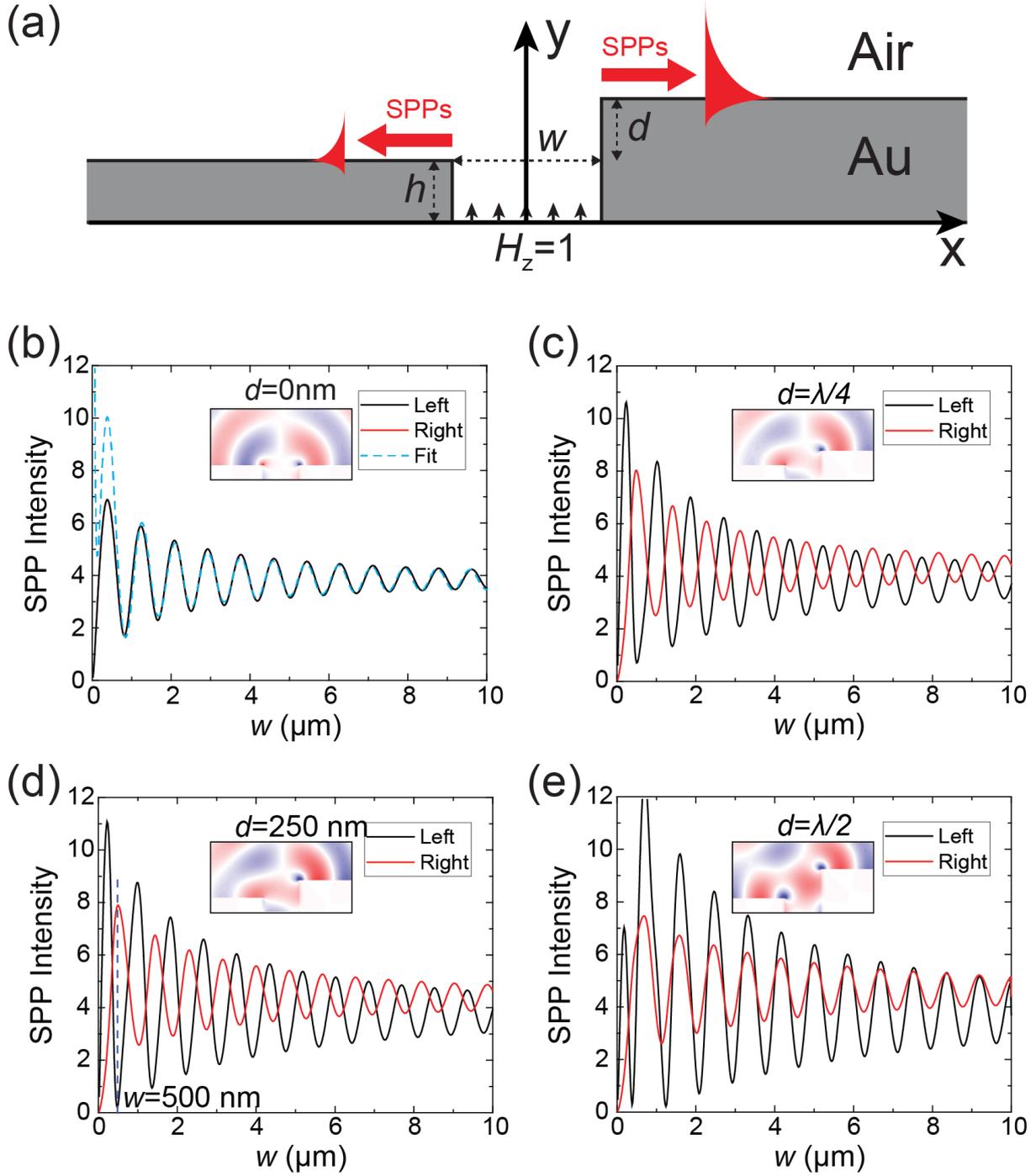

**Figure 1.** (a) Sketch of the metallic slit structure. Intensities of the right- and left-propagating SPPs generated by the step-slit structure as a function of the slit width $w$ at (b) $d = 0$ nm, (c) $d = \lambda/4$, (d) $d = 250$ nm, and (e) $d = \lambda/2$. The cyan dashed line in (b) denotes the fitted curve using Equation (4). The blue dashed line in (d) denotes the point of $w=500$ nm, which has a high extinction ratio. The insets in (b)-(e) show the electric field [Re($E_y$)] distributions at the slit width of $w=500$ nm.

To find the physics behind this phenomenon in the conventional symmetric slit as well as the step-slit structure, the field distributions of the structure are simulated and analyzed. Figure 2(a) and 2(b) show the magnetic field $[|\text{Re}(H_z)|^2]$ and electrical field $[|\text{Re}(E_y)|^2]$ distributions at the geometry of $w = 500$ nm and $d = 250$ nm where unidirectional SPP launching with a high extinction ratio is achieved, as denoted by the dashed line in Figure 1(d). It can be observed that the field is extremely strong around the corners, especially in Figure 2(b), where the strong electric fields at the corners form two hot spots evidently. The hot spots also emerge at the corners of the conventional symmetric slit of $d = 0$ nm.[33] The occurrence of the hot spots is attributed to the accumulated charges at the sharp corners which are illuminated by incident beams. Since the charges and currents are harmonic oscillating, these hot spots will radiate electromagnetic waves into the free space just like a harmonic dipole. The electromagnetic waves radiated from the hot spots can generate the SPPs when they impinge the plasmonic nanostructures on a metal film. Here, the electromagnetic wave radiated from the hot spots is termed as the hot spot cylindrical wave (HSCW). Different from the quasi-CW, which is a surface wave of a p-polarized state,[34–36] HSCW is a radiated wave in the free space and is independent of the polarizations.[29] As shown in Figure 2(c), the right- and left-propagating SPP amplitudes (denoted by $H^R$ and $H^L$, respectively) are composed of two parts: the pure-SPP parts ($H_{SP}^R$ and $H_{SP}^L$) and the hot-spot parts ($H_{HS}^R$ and $H_{HS}^L$). Herein, the pure-SPP parts refers to the SPPs that are generated purely from SPP modes, such as the waveguide modes in the lower-slit and the SPPs along the upper-slit. In the case of large $w$ ($w \gg \lambda$), we take the approximation that the SPPs in each metal wall of the silt are nearly independent, which means they don't couple with each other to construct a metal-isolator-metal (MIM) waveguide.[37] The complex SPP amplitudes at the right and left sides of the up-port of the lower-slit are denoted by $A_R$ and $A_L$, respectively. At the rectangular sharp corners, the SPPs along the metal wall of the slit can be scattered to the SPPs along the front metal surface, and the scattering coefficient is assumed to be *S*. Hence, the pure-SPP parts of the right- and left-propagating SPPs along the front metal surface can be expressed as

$$H_{SP}^{R} = A_{R} S \exp(i k_{SPP} d) \qquad (1a)$$

$$H_{SP}^{L} = A_{L} S \qquad (1b)$$

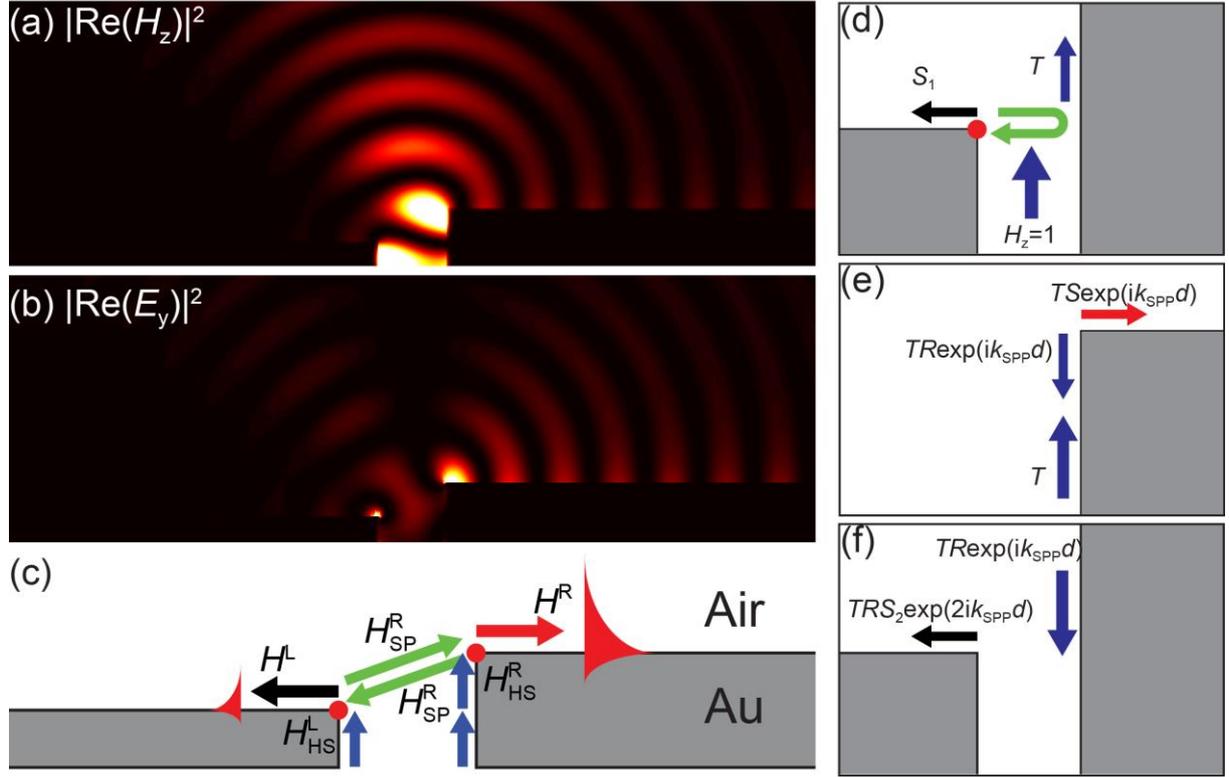

**Figure 2.** (a) Magnetic field and (b) Electrical field distributions in the step-slit structure with the geometric parameters of $w = 500$ nm and $d = 250$ nm. (c) Illustration of the different parts for the SPP amplitude. $H^R$ and $H^L$ are the total amplitudes of the SPPs propagating to the right (red arrow) and to the left (black arrow) of the slit. $H_{SP}^R$ and $H_{SP}^L$ are the pure-SPP parts contributing to the total SPP amplitudes, indicated by the blue arrows. $H_{HS}^R$ and $H_{HS}^L$ are the hot-spot parts contributing to the total SPP amplitudes, indicated by the green arrows. Illustrations of different stages of the semi-analytic model: (d) Electromagnetic fields in the lower-slit generate the left-propagating SPPs along the front metal surface and transmit to the SPPs along the upper-slit. (e) Upward propagating SPPs along the upper-slit partly generate the right-propagating SPPs along the front metal surface and are partly reflected at the top. (f)

Reflected SPPs along the upper-slit generate left-propagating SPPs along the front metal surface.

As for the hot-spot parts, it refers to the SPPs that are generated by the HSCW. Analog to the cylindrical wave radiated from the line sources into the free space, it's expected that the HSCW amplitude decays with the propagation distance in a negative power exponent, and its wave vector equals the wave vector in the free space. In addition, an approximation is made that the reflection of the HSCW by the opposite side of the slit is neglected. In other words, only the HSCW from one hot spot propagating to the opposite side is considered. This approximation is valid if the two hot spots are far away from each other ($> 3\lambda$). Hence, the hot-spot part can be written as

$$H_{HS}^{R} = \frac{B_L S_R'}{\left(\sqrt{d^2+w^2}\right)^m} \exp\left(ik_0\sqrt{d^2+w^2}\right) \quad (2a)$$

$$H_{HS}^{L} = \frac{B_R S_L' \exp(ik_{SPP}d)}{\left(\sqrt{d^2+w^2}\right)^m} \exp\left(ik_0\sqrt{d^2+w^2}\right) \quad (2b)$$

Here, $B_L$ and $B_R$ are the complex amplitude of the hot spots on the left and right sides of the metallic slit. $m$ is the attenuation index of the HSCW in the free space. $S_R'$ and $S_L'$ are the scattering coefficient for the HSCW at the right and left corners converting to the SPPs along the front metal surface. The factor of $\exp(ik_{SPP}d)$ denotes that the SPPs ($A_R$) at the right up-port of the lower-slit propagate for a length of $d$ along the metal wall of the upper-slit, as shown in Figure 2(c).

The total SPP amplitudes are the sum of the pure-SPP parts and the hot-spot parts, and the SPP intensities are proportional to the square of the SPP amplitudes. Thereby, the SPP intensities of the right- and left-propagating SPPs can be approximately expressed as Equation (3a) and (3b), respectively.

$$I^R \propto |H_{SP}^R + H_{HS}^R|^2 = \left| A_R S\exp(ik_{SPP}d) + \frac{B_L S_R'}{\left(\sqrt{d^2+w^2}\right)^m} \exp\left(ik_0\sqrt{d^2+w^2}\right) \right|^2 \quad (3a)$$

$$I^L \propto |H_{SP}^L + H_{HS}^L|^2 = \left| A_L S + \frac{B_R S_L' \exp(ik_{SPP}d)}{\left(\sqrt{d^2+w^2}\right)^m} \exp\left(ik_0\sqrt{d^2+w^2}\right) \right|^2 \quad (3b)$$

Based on Equation (3a) and (3b), it's evident that the hot-spot parts can bring the oscillation pattern to the SPP intensity curves. Since the wave vector of the HSCW equals the wave vector of light in the free space, this oscillation period is supposed to be about $T = 2\pi/k_0 = \lambda = 830$ nm. This is in good agreement with the simulation results of the oscillation period [Figure 1(b-e)]. In the special case of the symmetric slit ($d = 0$), Equation (3a) and (3b) become Equation (4),

$$I^R = I^L = \left| A + \frac{B}{w^m} \exp(ik_0 w) \right|^2 \quad (4)$$

where $A$ and $B$ are constant coefficient to be fitted. The fitting curve by using Equation (4) is depicted by the cyan dashed line in Figure 1(b). It is observed that the fitting data with an attenuation index $m = 0.8$ shows a good agreement with the simulation results. However, it's noticed that for small $w$, the fitting curve doesn't match well with the simulation data. This is because the two hot spots will influence each other when they are too close, and thus the coefficient $B$ (denoting the amplitudes of the hot spots) is no longer constant. Besides, the fitted results show that the complex coefficients of $A$ and $B$ have a relative phase difference of about $\pi$. Based on Equation (4), the SPP intensity becomes a minimum when the slit width equals even multiple of the half wavelength while the SPP intensity reaches a maximum when the slit width equals odd multiple of the half wavelength, according well with the previous results.[31,32] The reason is that the charges at the corners of the symmetric slit exhibit opposite signs, as shown by the inset in Figure 1(b). This mechanism also explains the synchronous oscillation curves in the case of $d = \lambda/2$ [Figure 1(e)]. In Figure 1(e), it is noticed that the SPP intensity becomes a minimum when the slit width equals odd multiple of the half wavelength while the SPP intensity reaches a maximum when the slit width equals even multiple of the half wavelength. This is attributed to the same signs of the charges at the corners, as shown by the inset in Figure 1(e).

In the case of the asymmetric step-slit structure ($d > 0$), for large $w$ ($w \gg d$), Equation (3a) and (3b) can be simplified to be

$$I^R \propto \left| A_R S + \frac{B_L S'_R \exp(ik_0 w)}{w^m} \exp(-ik_{SPP} d) \right|^2 \quad (5a)$$

$$I^L \propto \left| A_L S + \frac{B_R S'_L \exp(ik_0 w)}{w^m} \exp(+ik_{SPP} d) \right|^2 \quad (5b)$$

According to Equation (5a) and (5b), when $d$ increases, the oscillation curve of the right-propagating SPPs moves to the right of the $x$-axis [Figure 1(b)-1(e)] with a phase shift of $-k_{SPP} d$, while the oscillation curve of the left-propagating SPPs moves to the left of the $x$-axis [Figure 1(b)-1(e)] with a phase shift of $+k_{SPP} d$. Thus, their relative phase difference is $\varphi = 2k_{SPP} d \approx 2k_0 d$. This is in good accordance with the above simulation results [Figure 1(b)-1(e)] and analysis.

In order to further validate the above analysis, a semi-analytic model is first proposed to compute the pure-SPP parts of the launched SPPs along the front metal surface, then the deviation between the pure-SPP parts and the total launched SPPs is analyzed. As shown in Figure 2(d)-2(f), the SPP launching process in the step-slit can be divided into three simple stages. In the first stage, as shown in Figure 2(d), the incident light first generates the SPP mode in the lower-slit. This SPP mode can generate the left-propagating SPPs along the front metal surface with a scattering coefficient of $S_1$, as denoted by the black arrow in Figure 2(d). This scattering coefficient equals $S$ in Equation (1a) and (1b) when $w$ is large. Meanwhile, the SPP mode can convert into the SPPs along the right metal wall of the upper-slit with a transmission coefficient of $T$, as depicted by the blue arrow in Figure 2(d). As the SPPs propagate along the right metal wall of the upper-slit, its complex amplitude becomes $T\exp(ik_{SPP}d)$ when it reaches the top-edge of the step-slit structure. Then, it will partly generate the right-propagating SPPs along the front metal surface [with the complex amplitude of $TS\exp(ik_{SPP}d)$, as shown by the red arrow in Figure 2(e)] and partly be reflected [with the complex amplitude of $TR\exp(ik_{SPP}d)$, as denoted by the blue down arrow in Figure 2(e)]. This is the second stage, as depicted by Figure 2(e). In the final stage, as depicted in Figure 2(f), the reflected downward propagating SPPs along the right metal wall of the upper-slit can generate the left-propagating SPPs along the front metal surface [with the complex amplitude of $TRS_2 \exp(2ik_{SPP}d)$], as denoted by black arrow in Figure 2(f). Since the reflection coefficient $R$ at the top of the upper-slit is quite small, multiple reflections of the SPPs along the right metal wall of the upper-slit are neglected. For each stage, the corresponding scattering coefficient ($S$, $S_1$, $S_2$), reflection coefficient ($R$), and transmission coefficient ($T$) can be calculated using the mode orthogonality condition,[32,35,36,38] and their values are only dependent on the slit width $w$. However, in the first stage, the right metal wall of the upperslit can greatly reflect the radiated field from the left hot spot if $w$ is small, as shown by the green arrow in Figure 2(d). Hence, the numerical value of $S_1$ also includes some hot-spot parts, and the value is dependent on both of $w$ and $d$. For simplification, the value of $S_1$ is numerical calculated when $d$ is infinite and $w$ is a fixed value. The influence of the simplification on the model will be discussed later. Here, the slit width is fixed to be $w = 500$ nm in consistent with our previous simulation geometry [Figure 2(a)]. Therefore, the pure-SPP parts of the SPP amplitudes in Equation (1a) and (1b) can be semi-analytically expressed as

$$H_{SP}^{R} = TS\exp(ik_{SPP}d) \qquad (6a)$$

$$H_{SP}^{L} = S_1 + TRS_2\exp(2ik_{SPP}d) \qquad (6b)$$

By comparing Equation (6a) and (6b) with Equation (1a) and (1b), it is easy to obtain that $A_R = T$ and $A_L S = S_1$. The additional item in Equation (6b) comes from the reflected SPPs along the upper-slit, and this part is neglected in Equaiton (1b) by considering that the scattering coefficient $S_2$ tends to zero when $w$ is large. When $w$ is very small (e.g. $w = 500$ nm), this additional item should be considered.

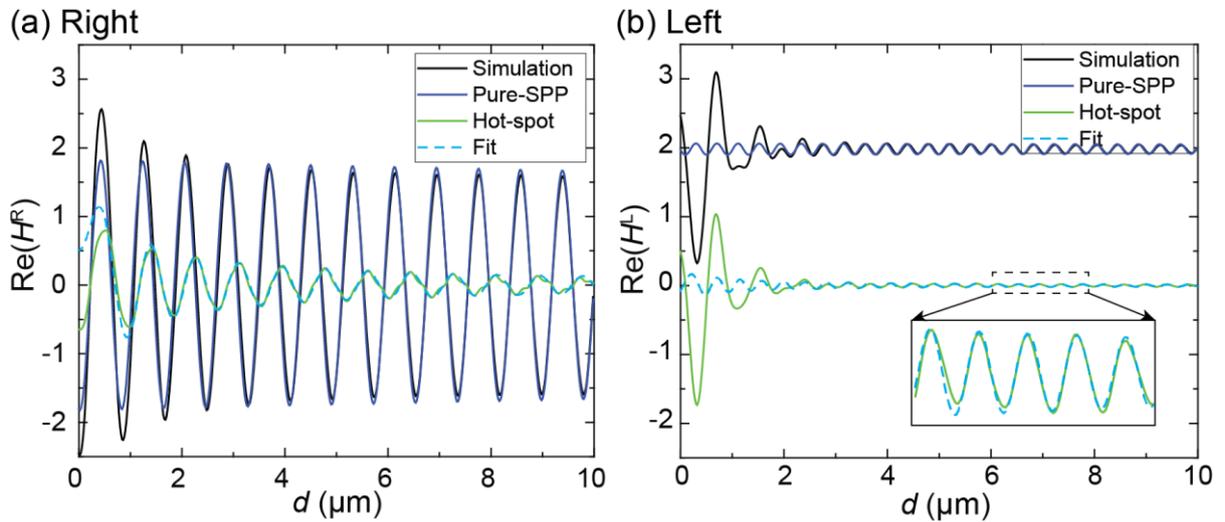

**Figure 3.** Amplitudes of the SPPs as a function of $d$ for $w = 500$ nm. Re($H$) for the (a) right- and (b) left-propagating SPPs. The black and blue lines denote the simulation results and the pure-SPP parts, respectively. The green lines denote the deviations between them. Cyan dashed lines denote the fitted curves using Equation (2a) and (2b). The inset in (b) shows a detailed view of the SPP amplitude between $d = 6$ μm and $d = 8$ μm.

The real part of the SPP amplitudes along the front metal surface as a function of $d$ is displayed in Figure 3. Here, the simulation results are denoted by the black lines, and the pure-SPP parts calculated using Equation (6a) and (6b) are denoted by the blue lines. The deviations between the simulation results and the pure-SPP parts are given by the green lines.

The fitted curves using Equation (2a) and (2b), which are the theoretically predicted values of the hot-spot parts, are depicted by the cyan dashed lines. For the right-propagating SPPs [Figure 3(a)], it's observed that the fitted curve using Equation (2a) (cyan dashed line) matches well with the deviation (green line) except for small $d$, as shown in Figure 3(a). This disagreement at small $d$ is attributed to that the two hot spots in Figure 2(c) will have interactions when they become too close to each other. The fitted attenuation index is $m = 0.8$, which is in good agreement with our previous fitting result [$m = 0.8$ in Equation (3a) and (3b)]. For the left-propagating SPPs [Figure 3(b)], it's seen that the fitted curve by using Equation (2b) (cyan dashed line) also matches well with the deviation (green line) for large $d$, as shown by the inset in Figure 3(b). As mentioned above, the reflection of the HSCW on the right metal wall of the step-slit structure has an influence on the left-propagating SPPs, and the scattering coefficient $S_1$ is calculated as the right metal wall goes infinite ($d\rightarrow\infty$). For large $d$, the reflection of the HSCW on the right metal wall is nearly constant, and the good agreement between the fitted curve (cyan dashed line) and the deviation (green line) is evidently observed in Figure 3(b). However, for small $d$, it's noticed that the mismatching of the left-propagating SPPs is much worse compared to that of the right-propagating SPPs. This is attributed to the different reflection behavior of the HSCW at the two sides of the step-slit structure. For the HSCW from the left hot spot, its right side is a metal wall. When the HSCW from the left hot spot impinges the metal wall, a large portion of its energy will be reflected back. Whereas for the HSCW from the right hot spot, most of its left side is air, and the reflection coefficient by the left side should be dramatically small. Therefore, the influence of the HSCW's reflection on the right-propagating SPP excitation is nearly negligible while the HSCW's reflection plays an important role in the left-propagating SPP excitation. Moreover, it's noticed that the amplitude of the fitted curve of the left-propagating SPPs from the right HSCW is also smaller than that of the right-propagating SPPs from the left HSCW. This is attributed to the anisotropic emission of the HSCW at the sharp corners. The HSCW will mainly radiate to the diagonal direction of the rectangular corner in the free space,[29] and thus the HSCW amplitude propagating from the right hot spot to the left corner of the step-slit structure is smaller than the HSCW amplitude propagating from the left hot spot to the right corner. To sum up, the SPP launching in the step-slit structure can be influenced by the HSCW, and thus the SPP launching can be manipulated with the modulation of the HSCW. As a result, unidirectional SPP launching with a high extinction ratio could be achieved by tuning the geometric parameters of the step-slit structure, as shown in Figure 2(a) and 2(b).

3. Experiments

In order to further validate the unidirectional SPP launching induced by the HSCW, a step-slit structure with a length of 6 μm is fabricated by focused ion beams in a 450-nm-thick gold film, which is evaporated on a glass substrate with a 30-nm-thick titanium adhesion layer. Figure 4(a) shows the scanning electron microscopy (SEM) picture of the experimental structure. Due to the limitation of the experiment technology, the left side of step-slit structure is a little rougher than the right side, as shown by the zoomed-in SEM image in Figure 4(b). Moreover, it is found that the deeper the left side of the step-slit is, the rougher its surface would become. Therefore, we performed our experiment at about $d = 150$ nm, and the other geometric parameters are measured to be about $w = 540$ nm and $h = 300$ nm. When a p-polarized white light (radius of about 100 μm) illuminates the step-slit structure from the back side, the SPPs are generated to propagate along the front metal surface. To scatter the confined SPPs, two decoupling gratings (period of 800 nm and separation of about 30 μm) are symmetrically fabricated on the two sides of the step-slit structure, as shown in Figure 4(a). The scattered light is collected by a long distance objective (100×, NA 0.5) and then imaged on a CCD. The CCD image is displayed in Figure 4(c). Evidently, the right decoupling grating is much brighter than the left one, indicating that the SPPs generated by the p-polarized white light mainly propagate to the right direction. To quantitatively measure the extinction ratios of the unidirectional SPP generation in such a broad bandwidth, the collected light is coupled to a fiber, which connects a spectrograph (Andor). The measured data are depicted in Figure 4(d). Here, the normalized intensity is defined as the quotient of the right- (or left-) propagating SPP intensity and the total SPP intensity of both sides. The extinction ratio of the unidirectional SPP launching can be obtained by calculating the quotient between the normalized intensities of the right- and left-propagating SPPs. As shown in Figure 4(d), it's observed that the experiment result matches well with the simulation data, revealing a unidirectional SPP launching which confirms our analysis. Since the roughness on the left side of step-slit structure [Figure 4(c)] will bring additional loss to the left-propagating SPPs, the experiment extinction ratio is a little larger than the simulation result. By using the HSCW, the SPP launching is controlled. This unidirectional SPP launcher with the lateral dimension of only 540 nm may find important application in plasmonic circuits.

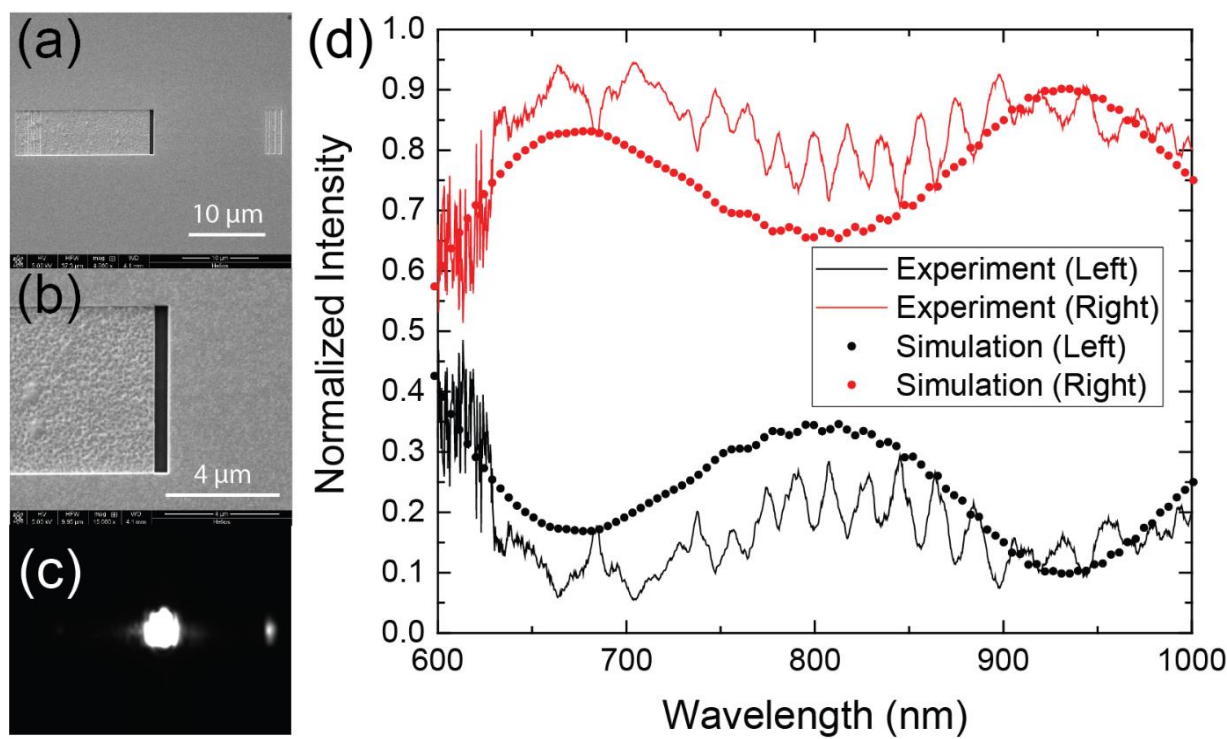

**Figure 4.** Experiment results. (a) SEM images of the experimental sample. (b) Zoomed-in SEM image of the step-slit structure. (c) CCD images when white light illuminates the sample

from the back side. (d) Wavelength responses of the normalized intensity. Normalized intensities of the right- and left- propagating SPPs are denoted by the red and black colors. The simulation and experiment results are denoted by dots and lines.

## 4. Conclusions

In summary, the influence of the hot spots on the propagating SPPs was theoretically and experimentally studied in the metallic slit structure. The simulation and semi-analytic model were implemented to explain the underlying physics of this phenomenon. It was found that the electromagnetic field radiated from the hot spots, termed as HSCW, could convert into the SPPs by scattering. This conversion greatly manipulated the SPP launching. As a result, unidirectional SPP launching in a broad bandwidth was experimentally realized in an ultra-small metallic step-slit structure by using the HSCW. The HSCW could bridge the localized surface plasmons and the propagating surface plasmons in an integrated platform, and thus it might provide a new degree of freedom in the design of SPP-based devices.

**Acknowledgements**
This work was supported by the National Basic Research Program of China (Grants 2013CB328704) and the National Natural Science Foundation of China (Grants 11204018, 61475005, and 11134001).
Received: ((will be filled in by the editorial staff))
Revised: ((will be filled in by the editorial staff))
Published online: ((will be filled in by the editorial staff))